\documentclass[reprint,nofootinbib,amsmath,amssymb,fontenc aps]{revtex4-1}
\usepackage{graphicx,dcolumn,bm,hyperref,float}
\usepackage{anyfontsize}

\newcommand{\be}{\begin{equation}}
\newcommand{\ee}{\end{equation}}
\newcommand{\ba}{\begin{eqnarray}}
\newcommand{\ea}{\end{eqnarray}}

\newcommand{\tc}{\kappa}
\newcommand{\mphi}{m_\phi}
\newcommand{\mt}{m_{\tiny{\mbox{top}}}}
\newcommand{\Mi}{M_i}

\newcommand{\Ni}{N_i}
\newcommand{\nni}{g_i}
\newcommand{\nh}{g_h}
\newcommand{\mpl}{M_{\tiny{\mbox{Pl}}}}
\newcommand{\Neff}{N_{\tiny{\mbox{eff}}}}
\newcommand{\Veff}{V_{\tiny{\mbox{eff}}}}
\newcommand{\Hinf}{H_{\tiny{\mbox{inf}}}}
\newcommand{\nphi}{n_\phi}
\newcommand{\fphi}{f_\phi}
\newcommand{\fequ}{f_{\tiny{\mbox{eq}}}}
\newcommand{\nequ}{n_{\tiny{\mbox{eq}}}}
\newcommand{\Treh}{T_{\tiny{\mbox{reh}}}}
\newcommand{\Hreh}{H_{\tiny{\mbox{reh}}}}
\newcommand{\treh}{t_{\tiny{\mbox{reh}}}}
\newcommand{\rhoSM}{\rho_{\tiny{\mbox{SM}}}}
\newcommand{\gb}{\gamma}
\newcommand{\Gann}{\Gamma_{\tiny{\mbox{ann}}}}
\newcommand{\dimc}{c}
\newcommand{\muF}{\mu_F}
\newcommand{\phia}{\phi_{\tiny{\mbox{amp}}}}

\begin{document}

\title{\fontsize{10.7}{15}\selectfont Explanation for why the Early Universe was Stable and Dominated by the Standard Model}

\author{Mark P.~Hertzberg$^*$}
\author{Mudit Jain$^\dagger$}
\affiliation{Institute of Cosmology, Dept.~of Physics and Astronomy, Tufts University, Medford, MA 02155, USA}

\begin{abstract}
The Standard Model (SM) possesses an instability at high scales that would be catastrophic during or just after inflation, and yet no new physics has been seen to alter this. Furthermore, modern developments in quantum gravity suggest that the SM degrees of freedom are not unique; that a typical low energy effective theory should include a large assortment of hidden sector degrees of freedom. It is therefore puzzling that cosmological constraints from BBN and CMB reveal that the early universe was almost entirely dominated by the SM, when the inflaton $\phi$ could have decayed into many sectors. In this work we propose the following explanation for all of this: we allow the lowest dimension operators with natural coefficients between the inflaton and both the Higgs and hidden sectors. Such hidden sectors are assumed to be entirely natural; this means all unprotected masses are pushed up to high scales and project out of the spectrum, while only massless (or protected) degrees of freedom remain, and so the inflaton can only reheat these sectors through higher dimension (and suppressed) operators. On the other hand, the SM possesses a special feature: it includes a light Higgs $H$, presumably for life to exist, and hence it allows a super-renormalizable coupling to the inflaton $\phi\, H^\dagger H$, which allows rapid decay into the SM. We show that this naturally (i) removes the instability in the Higgs potential both during and after inflation due to a tree-level effect that increases the value of the Higgs self-coupling from the IR to the UV when one passes the inflaton mass, (ii) explains why the SM is dominant in the early universe, and in particular we compute the relative temperature and abundances of the sectors, (iii) allows dark matter to form in hidden sector/s through subsequent strong dynamics (or axions, etc), (iv) allows for high reheating and possibly baryogenesis, and (v) accounts for why there so far has been no direct detection of dark matter or new physics beyond the SM.
\end{abstract}

\maketitle

%{\em Introduction}.---
\section{Introduction}
The Standard Model (SM) of particle physics and gravity has so far passed all experimental tests in the laboratory and solar system. Its structure has internal theoretical consistency with the laws of quantum mechanics and relativity, far beyond any theory previously established, such as the Fermi theory or the theory of elementary massive W-bosons. With the inclusion of the Higgs particle and the graviton, it provides for the first time a unitary theory up to the Planck scale, and it is an open question at what scale it first breaks down. 

On the other hand, from the top down point of view, there is currently no understanding why the SM has the particular set of degrees of freedom that it does. In fact our leading theory of quantum gravity, string theory, suggests there should be many {\em hidden sectors}; new degrees of freedom beyond the SM involving new gauge groups and various new types of particles (e.g., see \cite{Giedt:2000bi,Taylor:2015ppa}). Furthermore, astronomical tests reveal to us new physics beyond the SM, including dark matter, the need for a baryon asymmetry, and the need for early universe inflation. 

The existence of dark matter appears at first sight to be in accord with such top down expectations; namely one can imagine that the dark matter arises out of one or more of these hidden sectors. This is an idea that we will discuss more later in this paper. However, this idea of many hidden sectors seems to be in great tension with the following observational fact: Both big bang nucleosynthesis (BBN) and cosmic microwave background radiation (CMB) measurements tell us that the early universe was in fact almost entirely dominated by the SM light degrees of freedom (photons, neutrinos, etc). The current bound is that any new particles constituted $<4\%$ (at 95\% confidence) the total energy of the early universe. If there are in fact many hidden sectors, including those that provide dark matter, etc, this seems quite puzzling.

In terms of the SM's own internal consistency, there is in fact a sub-Planckian scale at which it may be breaking down, giving us the first need for new physics. As is well known, if we take the SM and run its coupling up to high scales using the known renormalization group (RG) equations, then there is a potential problem with the Higgs $H$'s self-coupling $\lambda$ (the coefficient of the quartic term $\Delta\mathcal{L}=-\lambda\,(H^\dagger H)^2$). At two-loops the running of $\lambda=\lambda(E)$ is given as the solid lines in Fig.~\ref{RunningBoth}. The coupling goes negative for energies $E\gtrsim 10^{10-11}$\,GeV, depending on the top mass. This means that the corresponding effective Higgs potential $\Veff(h)\approx\lambda(h)\,h^4/4$ turns over and goes negative at high scales too \cite{EWvevmetastbl.1,EWvevmetastbl.2,EWvevmetastbl.3,EWvevmetastbl.4}. 

It is known that this turn over in the potential renders the lifetime of our electroweak vacuum only meta-stable, with a lifetime that is much longer than the current age of the universe. However, this can have disastrous consequences during the early universe, especially during or just after cosmic inflation, since the Higgs field can exhibit fluctuations that take it to the unfavorable side of the potential \cite{Espinosa.Riotto.1,Zurek.1,Zurek.2}. During inflation, there are de Sitter fluctuations in the Higgs $\sim \Hinf/(2\pi)$, which places a bound on the inflationary Hubble scale of $\Hinf\lesssim 10^{9}$\,GeV, as we determined precisely in the context of eternal inflation recently \cite{Jain:2019wxo,Jain:2019gsq,Hertzberg:2018kyi}. Even if this condition is satisfied during inflation, there are still potential problems in the post-inflationary era; if there are non-negligible couplings between the inflaton and the Higgs, such as $\Delta\mathcal{L}=-\tc\,\phi\,H^\dagger H-\dimc\,\phi^2 H^\dagger H$, this can lead to large parametric resonance that can again cause the Higgs field to fluctuate to the unfavorable side \cite{Herranen:2015ima,Ema:2016kpf,Enqvist:2016mqj,Kohri.1,Kohri.2,Gross:2015bea}. The trilinear term $\sim \phi\, H^\dagger H$ seems especially dangerous because it will cause the Higgs field to become tachyonic as $\phi$ oscillates, leading to explosive fluctuations in $H$. 

The traditional solution to the above is to assume the Hubble scale of inflation is small and the coupling of the Higgs to the inflaton is small to avoid the post-inflationary disaster \cite{Figueroa:2015rqa,Joti:2017fwe,Ema:2016ehh,Markkanen:2018pdo,Saha:2016ozn,Ema:2017rkk,Gong:2017mwt}. However, low scale inflation arguably requires more fine-tuning of the inflationary potential, especially since the bound $\Hinf\lesssim 10^9$\,GeV has to be satisfied all the way back to very early times. Moreover, the need to have a small coupling between the inflaton and the Higgs only exacerbates the problem we mentioned earlier: it makes it difficult to efficiently reheat the visible SM sector, relative to many possible hidden sectors, and therefore we are left with the central puzzle as to why the early universe was dominated by the SM.

In this paper, we point out that there is a very reasonable framework that resolves all of these issues. We allow for naturally {\em large} couplings between the inflaton and the Higgs, including the trilinear coupling $\sim\phi\,H^\dagger H$, and point out that for large couplings, there is a tree-level correction to the Higgs potential that can remove the instability altogether for reasonable values of the inflaton mass, and for any scale of inflation. The regime in which this may or may not work is computed in detail in this work.

Furthermore, we explore the consequences of having hidden sectors with entirely {\em natural} parameters. The basic principle of naturalness suggests that if any unprotected (by a symmetry) masses are allowed for hidden sector particles, their masses should be taken to very high scales. In particular, this can be heavier than the inflaton, and therefore can project out of the physical spectrum relevant to the post-inflationary era. On the other hand, massless (or very light particles, such as axions with an approximate shift symmetry or Dirac fermions with an approximate chiral symmetry) can remain. This is all compatible with the idea of {\em technical naturalness}. 

Given this imposition of naturalness, one must then consider the status of the Higgs mass in the SM, which has a relatively light Higgs. One may wonder why it is light. In this case, however, we must bear in mind that the visible sector is fundamentally different than the hidden sectors. In particular, we are evidently built out of the SM particles, whose (fermionic) masses are proportional to the Higgs vacuum expectation value (VEV) $v$. Using this fact, it has been shown in Ref.~\cite{Agrawal:1997gf} that this demands the Higgs VEV $v$ is not too large in order for life to exist. (This argument applies for moderate to large changes in $v$ and the first generation masses $m_{u,d,e}$. However, if $v$ is increased by many orders of magnitude, then new phenomena can emerge \cite{Hall:2014dfa}). Associated with this, and recalling that the Higgs mass is $m_H=\sqrt{2\lambda}\,v$, one needs the Higgs mass to not be too large too (since large values of $\lambda$ are not allowed by unitarity). On the other hand, we are {\em not} built out of dark sectors, so this allows the dark sector masses to be entirely natural. Hence, given the need and existence of the SM Higgs, we have have a super-renormalizable coupling to the inflaton $\sim\phi\,H^\dagger H$, while all natural hidden sectors would primarily only couple to the inflaton through higher dimension operators $\sim\phi\, G_{i\mu\nu}G_i^{\mu\nu}$, etc. As we show, this naturally produces much colder hidden sectors, explaining why the SM is dominant in the early universe. Furthermore, dark matter can readily form from hidden sectors at later times due to strong coupling effects, as outlined by some of us in Ref.~\cite{Hertzberg:2019bvt}. We show that pure Yang-Mills hidden sectors can be in good agreement with galactic constraints due to its low temperature. (Alternatively, dark matter could arise from axions, etc.) We also comment on how this framework can potentially accommodate baryogenesis immediately after inflation and accounts for why we have yet to see any new physics beyond the SM.

%{\em Inflaton Couplings}.---
\section{Inflaton Couplings}
For a real scalar inflaton $\phi$, we write down all possible Lorentz invariant interactions. In the SM the Higgs is special in that it possesses a dimension 2 operator $H^\dagger H$, related to the Higgs mass term. For the hidden sector, we assume that all couplings are natural, hence all bare masses are assumed to be taken towards some high unification scale. This implies they effectively project out of the spectrum and are produced by the inflaton in negligible quantities. This leaves dimension 4 kinetic terms for massless (or nearly massless due to some symmetry) particles as the most relevant. The leading interactions with the SM and the hidden sector are therefore dimension 3 and dimension 5 operators, respectively
\be
\Delta\mathcal{L} = -\tc\,\phi\,H^\dagger H - \sum_i {\phi\over 4\Mi}\mbox{Tr}[G_{i\mu\nu} G_i^{\mu\nu}]+\ldots
\ee
where $G_i$ represents some hidden sector gauge bosons. We also anticipate dimension 5 couplings to hidden sector fermions $\sim\phi\,\bar{\psi}\,\slash\!\!\!\partial\,\psi$, but this does not lead to rapid decay of inflaton, due to helicity suppression, and so it is not as important. For vector-like fermions, a small Dirac mass may be possible by appealing to an approximate chiral symmetry; however, the dimension 4 Yukawa coupling $\sim y\,\phi\,\bar{\psi}\,\psi$ will then be small too, in accord with technical naturalness. One can check that this self-consistently makes the decay rate into light vector fermions very small. 
Note that we can also include the dimension 5 operator $\phi\,\mbox{Tr}[G_{i\mu\nu} \tilde{G}_i^{\mu\nu}]/F_i$, where $\tilde{G}$ is the dual field strength, whose perturbative decay is similar to the operator included. An additional feature is that for $F_i$ significantly lower than the Planck mass, then tachyonic resonance into $G_i$ can be possible \cite{Cuissa:2018oiw}. However, the resonance into the Higgs turns out to be much stronger. Also, by assuming an approximate CP symmetry, this term is in fact suppressed.

Hence, {\em by assuming that the dark sectors are natural, we will exploit in this work the idea that the inflaton can only couple to the dark sectors with higher dimension operators, but can couple to the Higgs with a low dimension operator.} This does of course assume a distinction between the sectors, but this is certainly not tautological: the point is that the SM may have an un-naturally light Higgs for life to exist, but the dark sectors need not have any un-naturally light scalars since we are evidently not built out of dark degrees of freedom and this has profound consequences. Put differently, the observation of needing a light Higgs VEV is ordinarily thought to be only relevant to prevent fast decays of nuclei \cite{Agrawal:1997gf}, but does not obviously have any bearing on the contents of the reheated sectors of the universe; however, we show that it does. In fact this point of view leads to non-trivial consequences, including the relative temperatures of the sectors, that we will compute in detail in this work.

The coefficient $\tc$ will be constrained shortly by the requirements of unitarity and cosmic stability, while the scale $\Mi$ represents some high mass scale, perhaps on the order of the Planck mass $\mpl$ ($\equiv1/\sqrt{8\pi G}$). More generally, we can lift these coefficients to be function of $\phi$, as $\tc\,\phi\to f(\phi)$, $\phi/\Mi\to f_i(\phi)$, which may be important during inflation when the inflaton has very large field values, perhaps of order $\mpl$ (note that we need $f_i(\phi)>-1$ to avoid the dark gauge bosons becoming ghostlike, which provides extra motivation for a high value of $M_i$). However in the post-inflationary era, as $\phi$ red-shifts, these leading terms will be of most importance. The inflaton and Higgs sector of the potential is 
\be
V = {m_\phi^2\over2}\phi^2+\kappa\,\phi(H^\dagger H-v^2/2)+\lambda(H^\dagger H-v^2/2)^2+\ldots
\label{potential}\ee
so we are expanding around $\langle H^\dagger H\rangle=v^2/2$ and $\langle\phi\rangle=0$. Furthermore, in the regime that we will work, there are no other new minima introduced (for $\kappa\gg m\sqrt{\lambda}$ this would be problematic, but the idea is that we will not work in such a regime). So it is a well behaved vacuum. 

%{\em Avoiding Instability}.---
\section{Avoiding Instability}\label{AI}
To avoid the instability problem we can correct the effective Higgs potential by exploiting the direct coupling between the Higgs and the inflaton. One possibility would be to use the quartic coupling $\Delta\mathcal{L}=-\dimc\,\phi^2 H^\dagger H$, which leads to a loop correction to the running of $\lambda$ of the order $\Delta\lambda\sim\dimc^2/(4\pi)^2$, requiring $\dimc\gtrsim\mathcal{O}(1)$ to cure the problem. However, this is highly problematic because this also leads to a correction to the inflaton self-coupling $\sim\lambda_\phi\,\phi^4$ of similar size. Since any reasonable model of inflation has extremely small self-interactions to give rise to small density fluctuations, this would require significant fine-tuning. (We will discuss later that this same problem does {\em not} afflict our proposed solution).

\begin{figure}[t]
\centering
\includegraphics[width=1\columnwidth]{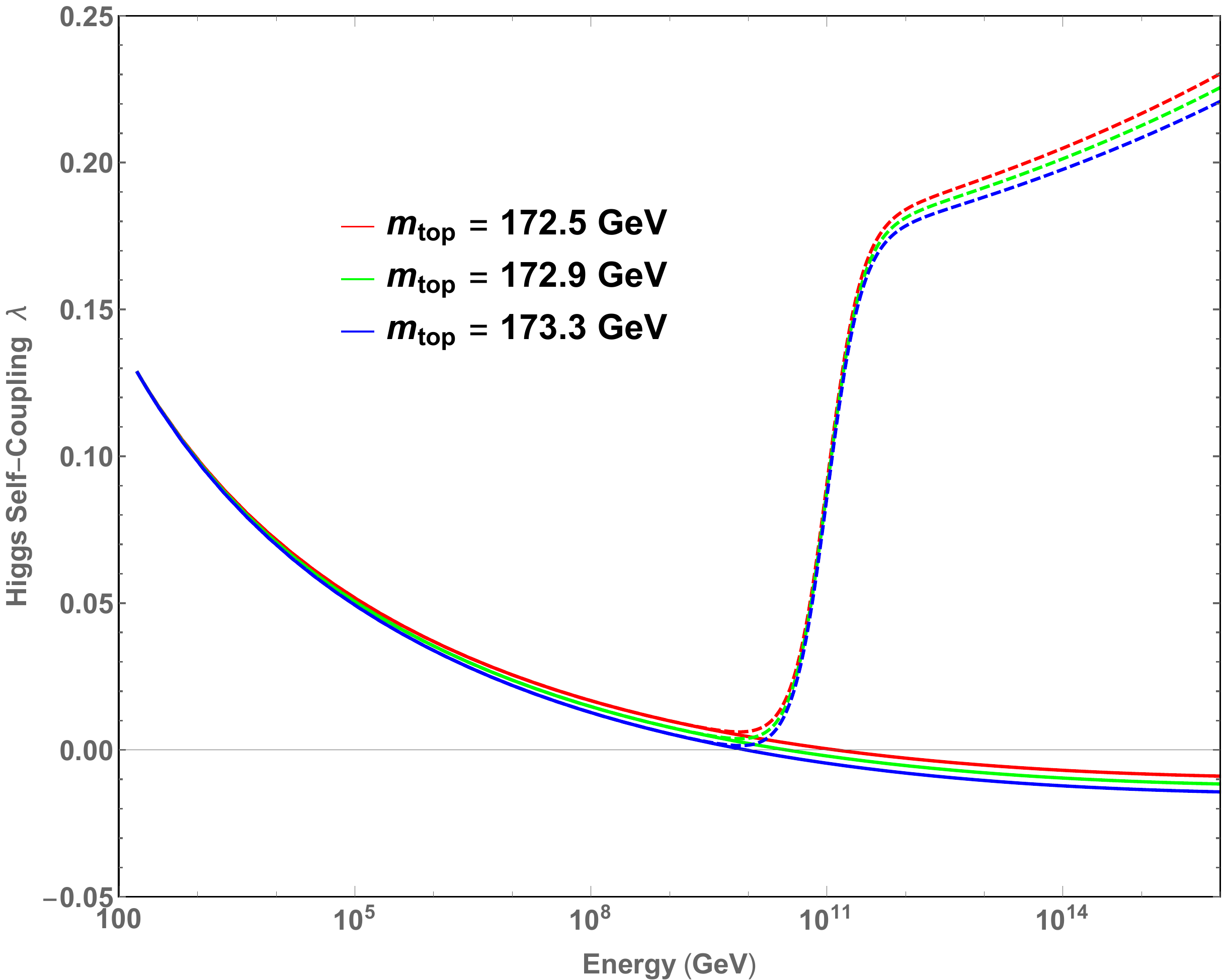}
\caption{Running of the Higgs self-coupling $\lambda$ versus energy (in units of GeV). The red curve is for $\mt=172.5$\,GeV, the green curve is for $\mt=172.9$\,GeV, and the blue curve is for $\mt=173.3$\,GeV. The solid curves are for the SM with no coupling to the inflaton, which runs negative, exhibiting an instability at high scales. The dashed curves are for the SM with a nonzero trilinear coupling to the inflaton $\Delta\mathcal{L} = - \tc\,\phi\, H^\dagger H$, with $\mphi=10^{11}$\,GeV and $\tc=0.6\,\mphi$, which no longer runs negative, removing the instability.}
\label{RunningBoth} 
\end{figure}

\begin{figure}[t]
\centering
\includegraphics[width=1\columnwidth]{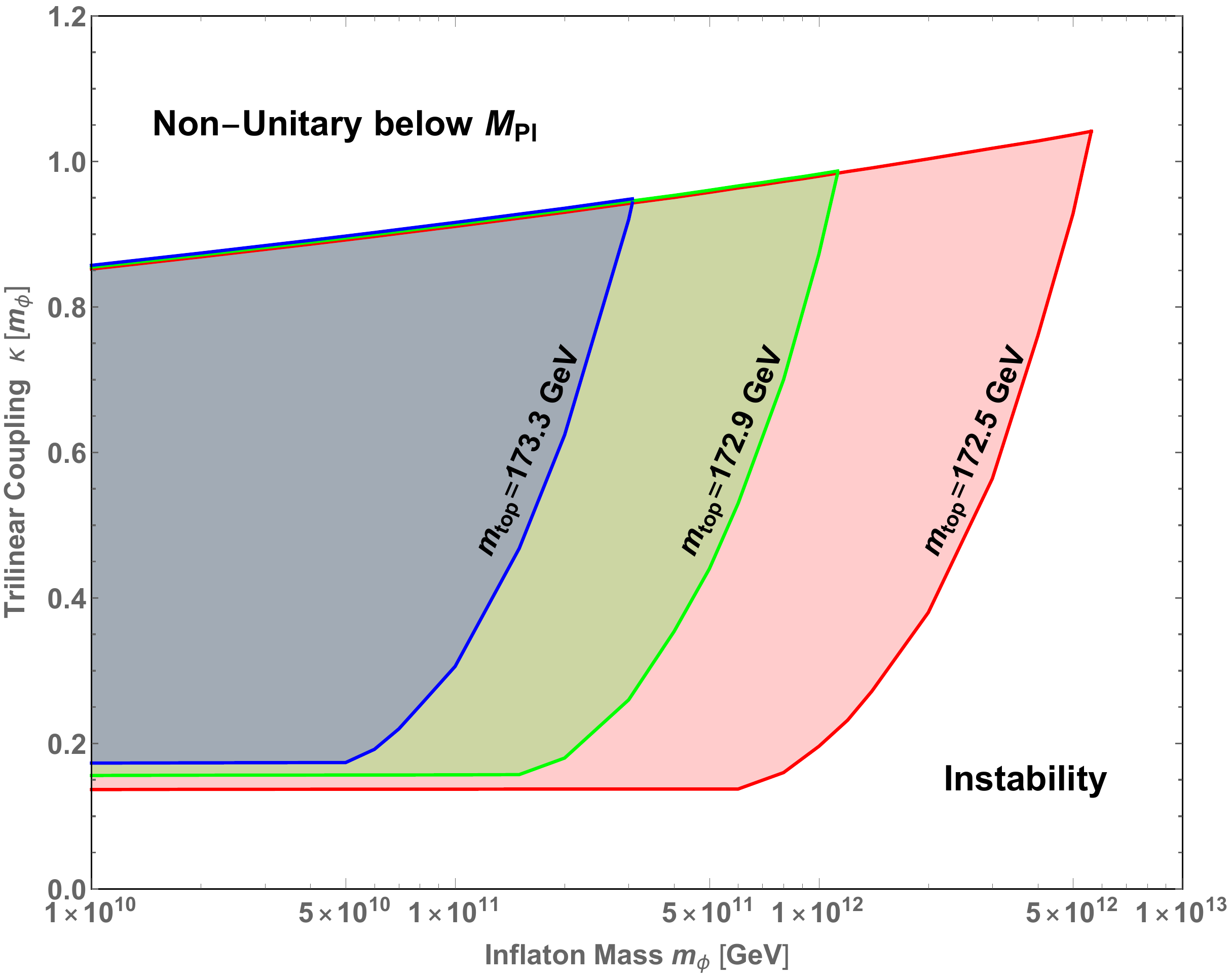}
\caption{Allowed region for the parameters in the model: the vertical axis is the trilinear coupling $\tc$ (in units of $\mphi$) and the horizontal axis is the inflaton mass $\mphi$ (in units of GeV). The red region is for $\mt=172.5$\,GeV, the green region is for $\mt=172.9$\,GeV, and the blue region is for $\mt=173.3$\,GeV (these are the central and 1 standard deviation uncertainties of the top mass \cite{PDB}). In the lower-right region the trilinear coupling $\tc$ is so small that the Higgs potential still exhibits an instability. In the upper-left region the trilinear coupling $\tc$ is so large that the theory violates unitarity below the Planck scale $\mpl$.}
\label{ConstraintPlot} 
\end{figure}

\begin{figure}[t]
\centering
\includegraphics[width=0.35\columnwidth]{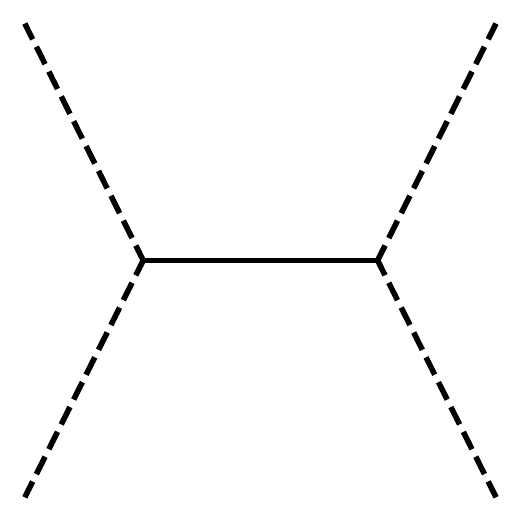}(a)\hspace{0.8cm}
\includegraphics[width=0.35\columnwidth]{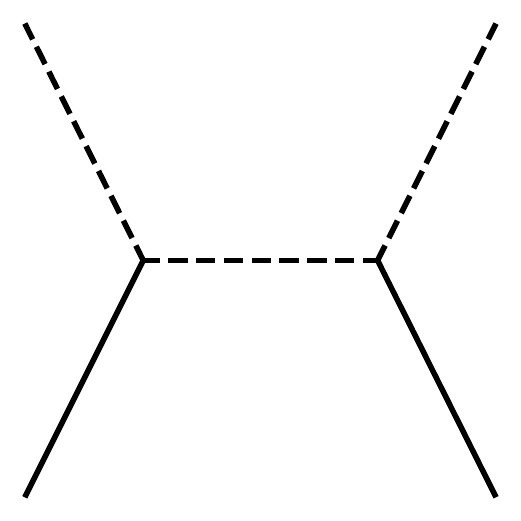}(b)\vspace{0.4cm}\\
\includegraphics[width=0.35\columnwidth]{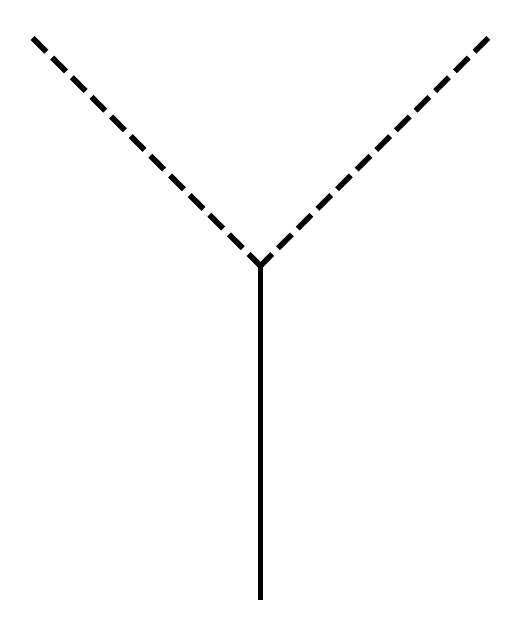}(c)\hspace{0.8cm}
\includegraphics[width=0.35\columnwidth]{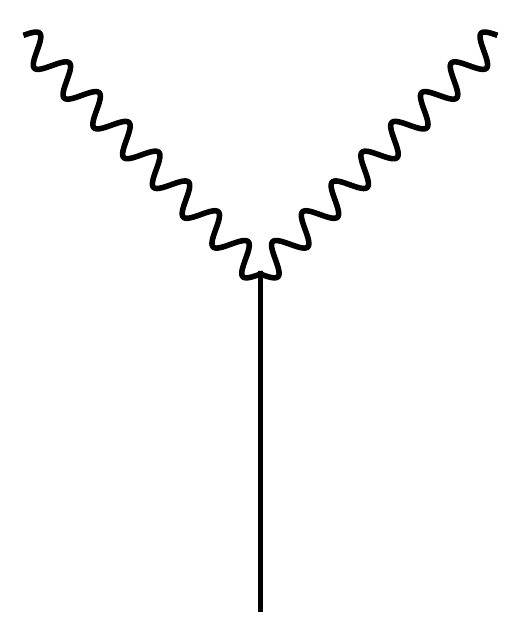}(d)
\caption{Some relevant processes in the paper. The dashed lines represent the Higgs, solid lines represent the inflaton, and wiggly lines represent hidden sector particles, such as hidden gauge bosons. (a) gives a tree-level correction to Higgs self-coupling $\lambda$, (b) \& (c) allow Higgs and inflaton to thermalize during reheating, and (d) allows inflaton to decay into hidden sector particles.}
\label{Diagrams} 
\end{figure}

In this work, we instead make use of the above trilinear interaction, which is much more promising. At tree-level the inflaton can mediate an exchange between a pair of Higgses. The $t$-channel diagram is given in Fig.~\ref{Diagrams}(a). By summing over all 3 channels, we have a contribution to the $hh\to hh$ scattering amplitude of
\be
\delta\mathcal{A}= \tc^2\left[{1\over s-\mphi^2}+{1\over t-\mphi^2}+{1\over u-\mphi^2}\right].
\ee
We evaluate this at $s=t=u=-E^2$, and it provides a negative contribution to $\lambda$ as we flow from the UV into the IR (attractive interaction). It is well known from the basic principles of effective field theory, that this leads to a tree-level shift in the effective $\lambda$ of the following (for a review, see Ref.~\cite{Burgess:2007pt})
\be
\lambda_{IR}=\lambda_{UV}-\kappa^2/(2\mphi^2)\,\,\,\,\,\,\,(\mbox{tree-level})
\label{TL}\ee
In order to study the Higgs effective potential in detail, it is important to include loop corrections. As a first approximation, one can incorporate this tree-level effect into the running of the self-coupling $\lambda$ by adding the following contribution to the beta-function
\be 
\beta_\lambda = {d\lambda\over d\ln E} = \beta_\lambda^{\tiny{\mbox{SM}}} + {\tc^2\, E^2\over(\mphi^2+E^2)^2},
\ee
where $\beta_\lambda^{\tiny{\mbox{SM}}}$ is the beta-function of $\lambda$ in the pure SM. Note that if one ignores the loops here (i.e., $\beta_\lambda^{\tiny{\mbox{SM}}}$), then one readily recovers Eq.~(\ref{TL}) as required.

Now the key idea is that by fixing $\lambda$ to its observed value at low energies $\lambda_{IR}\approx 0.13$, the above coupling to the inflaton provides a positive correction to the effective $\lambda$ as we flow to high energies $E\gtrsim \mphi$. Furthermore, note that this ensures that indeed no new additional minima are introduced, as we mentioned after Eq.~(\ref{potential}). We used the SM beta functions at 2-loops, solving for the evolution of Higgs self-coupling, the top quark Yukawa coupling, and the 3 gauge couplings. For sufficiently large values of $\tc$ and for sufficiently small values of $\mphi$, this can prevent the self-coupling and the effective potential going negative. An example of the effect on running is given by the dashed curves in Fig.~\ref{RunningBoth}. A similar idea with other kinds of (non-inflationary) scalars was presented in the important work of Ref.~\cite{EliasMiro:2012ay} (connections to axions was described in Ref.~\cite{Hertzberg:2012zc} and related ideas appears in Ref.~\cite{Gorbunov:2018llf})). One can readily confirm that in this regime the global minimum is at $\langle H^\dagger H\rangle=v^2/2$ and $\langle\phi\rangle=0$ and there are no other minima.

If the trilinear coupling $\tc$ is too large then the Higgs self-coupling increases to such large values that the theory violates unitarity at high scales. To determine the unitarity bound, the scattering amplitude $\mathcal{A}$ can be decomposed in partial waves $a_l$ as \cite{Krauss:2017xpj}
\be
\mathcal{A} = 16\pi\sum _l (2l+1)\,a_l \,P_l(\cos\theta).
\ee
For the $l=0$ partial wave, we have the unitarity bound Re$[a_0]\leq 1/2$. For $hh\to hh$ scalar scattering, we use the renormalized $\lambda(E)$ at some energy scale $E$, giving $\mathcal{A} = 6\,\lambda(E)$. Hence we have the unitarity bound
\be
\lambda(E)\leq {4\pi\over3}\,\,\,\,\,(\mbox{unitarity})
\ee
Since the self-coupling can grow with energy due to the trilinear coupling $\tc$, this puts an upper bound on $\tc$. For concreteness, we can demand that our theory remain unitary up to the Planck scale $\mpl=2.4\times10^{18}$\,GeV. In the upper-left region of Fig.~\ref{ConstraintPlot} this condition is violated.

On the other hand, if the trilinear coupling $\tc$ is too small, then the Higgs self-coupling still runs negative. This means the instability in the Higgs potential persists, which can have disastrous effects during or after inflation. In the lower-right region of Fig.~\ref{ConstraintPlot} this problem occurs.

In the colored regions of Fig.~\ref{ConstraintPlot} neither of the above problems occur. The allowed region is sensitive to the value of the top mass $\mt$. To make this figure we have fixed all other parameters to the central best-fit values: Higgs mass is $m_h=125.1$\,GeV, strong coupling is $\alpha_s(m_Z)=0.1181$, electromagnetic coupling is $\alpha(m_Z)=1/127.9$, and weak angle is $\sin^2\theta_w=0.23122$. For the top mass we have explored the favored region based on direct measurements \cite{PDB}, namely
\be
\mt = 172.9\pm 0.4\,\mbox{GeV}.
\ee
(the contribution from all lighter SM fermions is negligible), here the error bar is 1 standard deviation, though the uncertainty in the top mass is under debate \cite{Butenschoen:2016lpz}. 
We note that for the central value of the allowed top mass 172.9\,GeV, the inflaton mass can be as large as $\mphi\sim 10^{12}$\,GeV if we push $\tc$ towards its upper value allowed by unitarity $\tc\sim 1\,\mphi$. If the top mass is taken towards its lower value allowed by current data, then the inflaton mass can be even higher, including $\sim 10^{13}$\,GeV. The latter is near its preferred value in some of the simplest inflation models, such as chaotic inflation; in this model the potential is simply $V=\mphi^2\phi^2/2$ with $\mphi\approx 1.4\times 10^{13}$\,GeV. If we decrease the top mass from its central value by 1 standard deviation, as plotted as the red region in Fig.~\ref{ConstraintPlot}, we can have $\mphi$ as large as $5\times 10^{12}$\,GeV. While if we decrease the top mass by 2 standard devations, we can accommodate the chaotic inflaton mass of $1.4\times 10^{13}$\,GeV. 

 The reason inflation typically prefers these inflaton masses in the simplest models is that the ratio $~m/\mpl$ roughly sets the amplitude of primordial fluctuations, which is measured to be $\sqrt{\mathcal{P}}\sim 10^{-5}$. Although $\mphi$ can be smaller than this in other models. We do not claim to explain why $\sqrt{\mathcal{P}}\sim10^{-5}$, but we take this as a fact of nature, and may be explored in other work.

We note that since $\tc\lesssim\mphi$, the renormalization of the inflaton's mass $\Delta\mphi^2\sim\tc^2/(4\pi)^2$ is small, and so no additional fine-tuning is introduced in the inflaton sector. Furthermore, there are no quartic terms $\sim \lambda_\phi\phi^4$ generated in this model, as can be directly checked by computing the Coleman-Weinberg effective potential \cite{Coleman:1973jx}. 
In fact can easily complete the trilinear potential to have quartic terms $\sim \lambda_\phi\tilde\phi^4,\,\lambda_{\phi h}\tilde\phi^2 H^\dagger H$, as follows
\begin{eqnarray}
V=\lambda_\phi(\tilde\phi^2-\phi_0^2)^2+\lambda_{\phi h}(\tilde\phi^2-\phi_0^2)(H^\dagger H-v^2/2)\nonumber\\
+\lambda(H^\dagger H-v^2/2)^2
\end{eqnarray}
At low energies, we can expand around the inflaton's VEV as $\phi=\phi_0+\phi$, to obtain the inflaton mass as $m_\phi\sim \sqrt{\lambda_\phi}\,\phi_0$ and the trilinear coupling as $\kappa\sim \lambda_{\phi h}\,\phi_0$. For simple models of inflation, one usually has $\phi_0\sim \mpl\sim 10^{18}\,$GeV, so if $m_\phi\sim 10^{12}\,$GeV (see Fig.~\ref{ConstraintPlot}), we need $\lambda_\phi\sim 10^{-12}$ (which is well known for simple inflation models). If we also take $\kappa\lesssim m_\phi$ (again see Fig.~\ref{ConstraintPlot}), we need $\lambda_{\phi h}\sim 10^{-6}$. Since we already know that $\lambda\sim 0.1$ for the Higgs, this trio of quartic couplings ($\sim 10^{12},\,\sim 10^{-6},\,\sim 0.1$) is stable under loop corrections.  In particular, the cross term $\lambda_{\phi h}$ generates a 1-loop correction to both $\lambda_{\phi}$ and $\lambda$ of $\sim \lambda_{\phi h}^2/(4\pi)^2\sim 10^{-14}$, which is negligible compared to $\lambda$ and even smaller than $\lambda_\phi$. 
Indeed these all have self-consistently tiny corrections. (This is all in stark contrast to the simplistic solution mentioned at the start of this section, where there is no trilinear coupling, but only a quartic $\Delta\mathcal{L}=-c\,\phi^2H^\dagger H$ coupling, and using this at loop level to fix the instability. This demands $c\gtrsim \mathcal{O}(1)$ to cure the instability, leading to a much, much larger correction to $\lambda_\phi$.)

Returning again to the trilinear theory, there are of course loop corrections to the $2\phi\to2\phi$ scattering amplitude, but it can be readily checked that these effects are extremely small during inflation; the reason is that when we expand around the inflaton's large VEV (which is usually Planckian), the Higgs acquires a large effective mass $m_{eff}\sim \kappa\,\phi_0$, which significantly suppresses loop corrections. This ensures that we are studying a well behaved effective field theory, both during inflation, and at low energies after inflation.

%

%{\em Dark Radiation Relic Abundance}.---
\section{Dark Radiation Relic Abundance}
In the beginning of the post-inflationary era of ``preheating", the Higgs field may undergo parametric resonance as the inflaton oscillates coherently about its minimum \cite{Kofman:1997yn}. Since we are including the $\sim\tc\,\phi\,H^\dagger H$ interaction, this may be quite explosive. (In the literature, this resonance is often avoided by demanding that the (maximum) Floquet exponent $\muF\approx\tc\,\phia/(2\mphi)\lesssim$\,\,Hubble, which implies $\tc\lesssim\mphi^2/(\mpl)$. But this suppresses the decay into SM.) However, since there is no true instability remaining in our potential, a resonance is harmless; it quickly leads to fragmentation of the inflaton condensate, and then the system approaches more standard incoherent behavior.  

At this time, perturbative processes will become important, including the decay of the inflaton into Higgs particles and hidden sector gauge bosons $\gb_i$, etc, with rates (see Fig.~\ref{Diagrams} (c) \& (d))
\ba
\Gamma(\phi\to h\,h) &=&  {\nh\over32\pi}{\tc^2\over \mphi},\label{DecayHiggs}\\
\Gamma_i(\phi\to \gb_i \,\gb_i) &=& {\nni\over 128\pi}{\mphi^3\over \Mi^2},\label{DecayGluon}
\ea
where $\nh$ is the number of components of the Higgs, which in the SM is $\nh=4$, and $\nni$ is the number of dark boson helicities, e.g., for $SU(\Ni)$, we have $\nni=2(\Ni^2-1)$. Since we assume $\tc\sim\mphi$ (recall Fig.~\ref{ConstraintPlot}) and since $\Mi$ is assumed to be near some high scale, such as $\mpl$, we can be sure that the decay into the Higgs is fastest. The Higgs will then interact with itself and the rest of the SM and thermalize. This thermalization will be rapid and occur at the time $\treh\sim 1/\Hreh\sim 1/\Gamma(\phi\to h\,h)$. Using the Friedmann equation in a radiation dominated era $\Hreh^2=g_*\pi^2 \Treh^4/(90\mpl^2)$, gives the standard estimate for the reheat temperature $\Treh\approx 0.5\sqrt{\Gamma\,\mpl}$. For $\tc\sim\mphi$, we obtain 
\be
\Treh\sim 0.1\sqrt{\mphi\,\mpl}.
\ee
For the central top mass, we need $\mphi\lesssim 10^{12}$\,GeV, giving $\Treh\lesssim 10^{14}$\,GeV. This allows for such high reheat temperatures that new particles may be produced with renormalizable couplings to the SM. In particular, this may allow for baryogenesis in the visible sector to occur.
A possible example may be to make use of the dimension 5 Weinberg operator $\sim m_\nu(L\cdot H)^2/v^2$ to facilitate spontaneous baryogenesis. We note that this operator is suppressed by the scale $\Lambda=v^2/m_\nu\sim 10^{14}$\,GeV, which is higher than the scale at which the Higgs potential would turn over (in the absence of our trilinear coupling). So the coupling to the inflaton is still needed and is not significantly altered by this new term. Since we find a reheat temperature of $\Treh\sim 10^{14}$\,GeV, this seems to fit in nicely with the scale of the Weinberg operator, so it would be effective for at least a little while, before freezing out.

In order to determine the abundance of the hidden sectors, we need to determine the inflaton abundance and in turn the dark radiation abundances. The Boltzmann equation for the inflaton's occupancy number $\fphi(p)$ is
\be
\dot{f}_\phi(p)+3 H\fphi=\tilde{\Gamma}(p)(\fequ(p)-\fphi(p))+\ldots,
\ee
where $\tilde{\Gamma}(p)$ is the decay rate for a particle of momentum $p$. It is related to the ordinary decay rate $\Gamma$ by $\tilde{\Gamma}(p)=\Gamma\,\chi(p,T)/\gamma(p)$, where $\gamma(p)=m_\phi/E_\phi(p)$ is the Lorentz factor from time dilation and $\chi(p,T)\geq1$ is a factor that accounts for Bose enhancement. We are including both the forward and backward processes of 1 inflaton into 2 Higgses (since coupling to dark radiation is suppressed), as depicted in Fig.~\ref{Diagrams} (c), while the ``+\ldots" indicates other processes, such as 2 inflatons into 2 Higgses, as depicted in Fig.~\ref{Diagrams} (b), etc. It is easy to check that these rates are much larger than Hubble, ensuring thermal equilibrium of the inflaton with the SM for all temperatures. This implies that for $T\ll\mphi$ its abundance becomes Boltzmann suppressed $\nphi\propto e^{-\mphi/T}$.

On the other hand, the dark sectors are expected to never be in thermal equilibrium with the SM or the inflaton. For $\Mi$ near the Planck scale, it is clear that the production rate (from decays or annihilations) of dark radiation is always much smaller than Hubble. The decay rate was given earlier in Eq.~(\ref{DecayGluon}). Since it is given by 2 inverse powers of $\Mi$, while Hubble is only given by 1 inverse power on $\mpl$, it readily follows that it is smaller in the regime of interest $T\gtrsim\mphi$ (before the inflaton abundance plummets exponentially). Production from annihilations $\Gann=\nequ\langle\sigma v\rangle$ can be estimated (for $T\gtrsim\mphi$) as $\Gann(h\,h\to \gb_i\,\gb_i)/H\sim\tc^2\mpl/(T\Mi^2)$ and $\Gann(\phi\,\phi\to \gb_i\,\gb_i)/H\sim T^3\mpl/\Mi^4$. The former is maximal at $T\sim\mphi$ and is moderately sub-dominant to the decay process $\phi\to \gb_i\,\gb_i$ (since $\tc\lesssim\mphi$), while the latter is maximal at high temperatures. If we push $\Mi$ to be somewhat close to $\mpl$ and $T\sim \Treh$, then this ratio is parametrically $\sim (\mphi/\mpl)^{3/2}$, which is extremely small. 

Although it is out of equilibrium, the hidden sector will be produced by inflaton decay. We would now like to determine the final energy density of dark radiation. Since the inflaton is in thermal equilibrium with the SM, and its number density is changing over time, we can determine the final energy density of dark radiation produced from decays as
\be
\rho_i(t) = {\mphi\,\Gamma_i\over a^4(t)}\!\int_0^t dt'\,n_\phi(t')\,a^4(t'),
\ee
where $n_\phi$ is the number density of inflatons (we can ignore any Bose enhancement factors here, because the injection energy $\sim\mphi$ turns out to be much higher than the dark temperature). We switch the time variable to temperature, using $dt'=dT' (dt'/dT')=-dT'/(T'H')$ (using $a\propto 1/T$). We use the Bose-Einstein distribution $n_\phi \sim \int d^3k/(e^{E_k/T}-1)$ and the Friedmann equation to obtain the following result for the ratio of energy densities of dark species to the SM thermal bath at late times
\be
{\rho_i\over\rhoSM} = {5\sqrt{5}\,\pi^2\over7\sqrt{2}\,g_*^{3/2}}{\Gamma_i\,\mpl\over\mphi^2}.
\ee
Using the SM value for the number of relativistic degrees of freedom $g_*=106.75$ and the decay rate in Eq.~(\ref{DecayGluon}), we obtain the relative amount of total dark radiation as
\be
{\rho_d\over\rhoSM} \approx 2.5\times 10^{-5}\sum_i\nni{\mphi\mpl\over\Mi^2}.
\label{densityratio}\ee
We see that if $\Mi$ is pushed to some high scale (such as Planck scale, or even a little lower), we have $\rho_d\ll\rhoSM$, even if the number of degrees of freedom in the hidden sectors is huge. For illustration, if we take $\Mi=10^{17}$\,GeV, $\mphi=10^{12}$\,GeV, then we can still have $\mathcal{O}(10^5)$ dark degrees of freedom and still maintain $\rho_d/\rhoSM<0.001$ at very early times. 

Also, if we let $g_{i*}$ be the number of effective degrees of freedom in $i^{th}$ dark sector ($g_{i*}=g_i+7n_i/8$, where $g_i$ is number of light bosons and $n_i$ is number of light fermions), then we determine the ratio of temperatures of the $i^{th}$ dark sector to visible sector to be
\be
\xi_i\equiv{T_i\over T}\approx 0.006\left(g_i\,\mphi\over g_{i*}\,10^{12}\,\mbox{GeV}\right)^{\!1/4} \! \left(\mpl\over\Mi\right)^{\!1/2}.
\label{tempratio}\ee

%{\em Constraint From BBN and CMB}.---
\section{Constraint From BBN and CMB}
To compare this to constraints from BBN and CMB, we need to track the evolution down to temperatures $T\sim 1$\,MeV and below. If there is strong dynamics, some (or most) of the dark degrees of freedom can confine and convert their entropy into the remaining massless degrees of freedom. We call the number of degrees of freedom $\tilde{g}_{i*}$ at BBN after any possible confinement, and for the (dominant) visible sector we use notation $g_*(=106.75)$ and $\tilde{g}_*(=10.75$).

As is well known, at the time of BBN, the SM prediction for the effective number of neutrino species is $\Neff\approx3.046$ (slightly larger than 3 due to a small coupling during $e^+e^-$ annihilation). Hidden sectors contribute the additional amount \cite{Hertzberg:2019bvt}
\be
\Delta \Neff = {4\over7} \sum_i \xi_i^4\,\tilde{g}_{i*}\!\left(g_{i*}\,\tilde{g}_{*}\over \tilde{g}_{i*}\,g_{*}\right)^{\!4/3},
\ee
The current bound on the number of additional massless species is \cite{Aghanim:2018eyx} $\Delta \Neff < 0.30$ (95\% confidence), which translates into the bound on the additional energy density of $\Delta\rho/\rhoSM<4\%$ (95\% confidence), as mentioned in the introduction. By using the result in Eq.~(\ref{tempratio}), we have $\xi_i\sim 10^{-2}$, and so the BBN bound is readily satisfied even for a very large number of particles in hidden sectors.

%{\em Discussion: (i) Dark Matter}.---
\section{Discussion}
{\em (i) Dark Matter}.---
Although the temperatures and hence energy densities of the dark sectors are small at early times, this framework still readily allows for the formation of dark matter. If one or more dark sectors include chiral fermions and undergo confinement, then massive degrees of freedom can emerge, such as dark baryons etc, as outlined by some of us in Ref.~\cite{Hertzberg:2019bvt}. This can lead to dark matter as a thermal relic, with abundance
\be
\Omega_d \approx \sum_i \xi_i {0.26\over (18\,\mbox{TeV})^2\langle\sigma_i v\rangle}.
\ee
So although it is suppressed by a factor of the ratio of temperatures $\xi_i\sim 10^{-2}$, or so, this can still readily produce the observed dark matter abundance $\Omega_d\approx 0.26$, by either (a) exhibiting strong dynamics at the scale $\Lambda_i \sim 10^{1-2}\,\mbox{TeV}/\sqrt{\xi_i}$, or (b) compensating by exploiting many sectors. 

For pure Yang-Mills hidden sectors, glueballs can form with relic abundance $\Omega_i\sim\xi_i^3\,(\Lambda_i/(10\,\mbox{eV}))$ for $SU(N_i)$ with $N_i\sim 3$ \cite{Boddy:2014yra}. If one had $\xi_i\sim 1$, then this would be highly problematic, since one would need $\Lambda_i\lesssim 3$\,eV to avoid over closure, but this leads to huge scattering cross sections $\sigma_{\tiny{\mbox{sc}}}\propto 1/\Lambda^2$ in galaxies, which is clearly ruled out. However, if we take our result from 
Eq.~(\ref{tempratio}) with $\xi_i\approx 0.006$, we can have $\Lambda_i\sim 50$\,MeV, which is marginally compatible with bullet cluster constraints \cite{Yamanaka:2019aeq,Yamanaka:2019yek}, suggesting interesting deviations from CDM.

Other possibilities for dark matter include axions through the misalignment mechanism. Depending on parameters, hidden sector axions could over close the universe, so we must forbid such parameters in this scenario.

{\em (ii) Isocurvature}.---We note that in this framework, since both the visible and dark sectors all arise from the decay of the same inflaton $\phi$, the model predicts that the primordial fluctuations are adiabatic. This is in accord with all current observations \cite{Aghanim:2018eyx}. If the dark matter arises from axions, then non-trivial bounds may apply.

{\em (iii) Lightness of Higgs}.---In this framework the lightness of the Higgs is accepted; we do not require a direct dynamical resolution to the hierarchy problem. Instead it is assumed that the smallness of the electroweak scale (and hence the Higgs mass) is due to environmental selection effects; namely, if the electroweak scale were significantly larger, then all nuclei would decay \cite{Agrawal:1997gf}, and observers would not be present. This explanation for the lightness of the electroweak scale has been criticized (e.g., \cite{Strasslerblog}) by noting that an alternative for the smallness of the electroweak scale could have been provided by technicolor. This appears to be just as conducive to observers and would occur much more readily in some landscape scenario, since technicolor does not appear to suffer from any fine-tuning. 

However, our framework provides a new perspective on this issue; the Higgs provides a unique opportunity for the inflaton to decay  into the SM by the dimension 3 operator $\sim\phi\,H^\dagger H$, thus predominantly populating the visible sector (also see Ref.~\cite{Arkani-Hamed:2016rle}). On the other hand, this opportunity would be lost in a Higgs-less model like technicolor. Hence it is plausible that in any universe with technicolor, while the electroweak scale may be naturally small, there would be a relatively huge abundance of dark radiation. This leads to its own environmental problems by wiping out small scale structure \cite{Takahashi:2019ypv}, but this problem is avoided in a universe with a light Higgs.

{\em (iv) Inflaton mass and top mass}.---We can compare our findings for the preferred inflaton mass Fig.~\ref{ConstraintPlot}, to expectations from inflationary predictions. The squared-amplitude of fluctuations from inflation is known to be $\mathcal{P}\approx \Hinf^2/(8\pi^2\mpl^2\epsilon)$, with measured value $\mathcal{P}\approx 2\times 10^{-9}$. In the simplest models of chaotic inflation, $V= \mphi^2\phi^2/2$, this requires an inflaton mass of $\mphi\approx1.4\times 10^{13}$\,GeV, which is slightly high compared to what is allowed in Fig.~\ref{ConstraintPlot} (although it is allowed if we lower the top mass by 2 standard deviations). However, such models predict a large tensor-to-scalar ratio, so lower values of $\Hinf$ are required by constraints on primordial B-modes. In simple models this is (though not always) correlated with lower values of the inflaton mass. This is then nicely compatible with the required values from Fig.~\ref{ConstraintPlot} in order to stabilize the Higgs potential. 

We note that if the top mass were considerably higher, then very low inflaton masses would be required to avoid the instability, which seems unsatisfying. While if the top mass were considerably lower, then we would not want a trilinear coupling $\tc$ that were so large, because it would exacerbate the unitarity problems. Hence in some sense, the observed value of top mass is optimal in this scenario.

{\em (v) Lack of new physics at low energies}.---In summary, we have found an extremely minimal scenario that appears to accommodate all of the central features of our universe; vacuum and Higgs stability, the domination of the SM in the early universe, inflation, dark matter, and possibly baryogenesis too by allowing for high reheating. It does so by positing only that our sector is atypical, presumably for life to exist, giving us a light Higgs, while all the other hidden sectors are entirely natural. This prevents large direct couplings of the inflaton to light particles in hidden sectors and it suggests that new physics may only enter at very high scales associated with inflation. So this framework accounts for why there has so far not been any direct detection of dark matter or new physics beyond the SM in colliders or precision tests.

{\em Acknowledgments}.---We thank Prateek Agrawal, Patrick Fitzpatrick, Matt Kleban, Liam McAllister, Fabrizio Rompineve, McCullen Sandora, Wati Taylor, and Jesse Thaler for useful discussions. M.~P.~H. is supported in part by National Science Foundation Grants No. PHY-1720332, PHY-2013953.

$^*$mark.hertzberg@tufts.edu, $^\dagger$mudit.jain@tufts.edu

\end{document}